\documentclass[superscriptaddress,aps,prl,10pt,twocolumn,floatfix,showpacs,nofootinbib,longbibliography]{revtex4-1}
\usepackage{amsmath,amssymb,amsthm}
\usepackage{easybmat}
\usepackage[colorlinks=true,citecolor=blue,urlcolor=blue]{hyperref}
\usepackage[pdftex]{graphicx}
\usepackage{float}
\usepackage{times,txfonts}
\usepackage{braket}
\usepackage{color}
\usepackage{latexsym}
\usepackage{dcolumn}
\usepackage{epsf}
\usepackage{natbib}
\newcommand{\beginsupplement}{%
        \setcounter{table}{0}
        \renewcommand{\thetable}{S\arabic{table}}%
        \setcounter{figure}{0}
        \renewcommand{\thefigure}{S\arabic{figure}}%
     }

\begin{document}
\title{Anharmonicity can enhance the performance of quantum refrigerators}  
\author{Sourav Karar}
\email{sourav.karar91@gmail.com}
\affiliation{Department of Physics, Government General Degree College, Muragachha, Nadia 741154, India}
\affiliation{S. N. Bose National Centre for Basic Sciences, Block JD, Sector III, Salt Lake, Kolkata 700 106, India}
\author{Shounak Datta}
\email{shounak.datta@bose.res.in}
\affiliation{S. N. Bose National Centre for Basic Sciences, Block JD, Sector III, Salt Lake, Kolkata 700 106, India}
\author{Sibasish Ghosh}
\email{sibasish@imsc.res.in}
\affiliation{Optics and Quantum Information Group, The Institute of Mathematical Sciences, HBNI, C. I. T. Campus, Taramani, Chennai 600113, India}
\author{A. S. Majumdar}
\email{archan@bose.res.in}
\affiliation{S. N. Bose National Centre for Basic Sciences, Block JD, Sector III, Salt Lake, Kolkata 700 106, India}
\begin{abstract}
We explore a thermodynamical effect of anharmonicity in quantum mechanical oscillators. We show  that small quartic perturbations to the oscillator potential lead to an enhancement of performance of quantum refrigerators for both the Otto and Stirling cycles. A similar nonlinearity driven enhancement of performance is also observed for an analogous spin-qubit model of quantum refrigerators. We further demonstrate the robustness of improvement of the coefficient of performance versus the energy cost for creating anharmonicity. It is shown that the anharmonicity driven improvement in  performance is a generic effect at the quantum level for the experimentally realizable Otto refrigerator.
\end{abstract}

\maketitle 

{\it Introduction:-}
Quantum thermodynamics has attracted an upsurge of interest in recent years revealing certain novel features and generalizations over its classical counterpart~\citep{r1,quan1,quan2}. Various forms of the second law of thermodynamics~\citep{scully,cq1,brandao}, and linkages with resource theories of quantum coherence~\citep{LJR15,SAP17} are under active development due to their foundational perspectives as well as potential applications in real systems. Microscopic configuration with restricted degrees of freedom enables a system to exhibit quantum mechanical supremacy~\citep{QT} beyond the limits set by classical thermodynamics~\citep{cq1,cq2}. Quantum behaviour of working media, such as quantum harmonic oscillators~\citep{ho1,ho2}, two-level~\citep{tl1,tl2} and multi-level spin systems~\citep{ml1,ml2} is inherently connected to the figures of merit belonging to various thermodynamic cycles. 

Quantum heat engines or refrigerators~\citep{Alicki,Petersonetal} are appropriate test grounds for quantum thermodynamics, having potential applications in diverse areas such as nanotechnology~\citep{nano1,nano2} and information processing~\citep{goold,misra}. Single-mode bosonic (or spin-$\frac{1}{2}$) systems are widely used as working substances for quantum heat engines~\citep{ho1,ho2,tl1,quan1,quan2,harm}. Single mode harmonic oscillators have experimental realizations in trapped ions~\citep{TI1} and optomechanical systems~\citep{OM1}. However, implementation of the ideal harmonic oscillator is quite difficult in practice. On the other hand, since no realistic oscillator is perfectly harmonic, a small quartic perturbation term can be introduced in the potential~\citep{review, Anharmonic, Anharmonic1}, leading to analytical expressions for energy eigenvalues, valid for a few orders of the perturbation strength. Such anharmonic oscillators are experimentally realizable~\citep{Exp1,Exp2,Exp3,Exp4,Exp5,Exp6,Exp7,Exp8}. 

Non-linear perturbations in the arena of quantum optical set-ups~\citep{TI2,OM2} have been studied to investigate various kinds of non-classical effects~\citep{nonlin3,nonlin4}. Interesting proposals for generating and stabilizing quantum entanglement aided with non-linearity have been formulated~\citep{nonlin1,nonlin2}. In the present work we are motivated to investigate the impact of non-linearity on thermodynamic processes. To this end here we specifically consider the quantum Otto~\citep{quan1} and Stirling refrigerators~\citep{stirling}. The Otto and Stirling engines are prototypical thermodynamic cycles extensively studied in the literature~\citep{KR17} with recent progress in experimental implementation at the quantum level~\citep{BB12,RDT+16,Petersonetal}.

Our approach here is to employ first such an implementable anharmonic oscillator with quartic correction to the potential~\citep{Anharmonic,Anharmonic1}. We find that the co-efficient of performance of the two refrigerators, {\it i.e.,} the Otto  and Stirling refrigerators, are  enhanced through increased non-linearity in the form of larger strength of anharmonicity.  We further construct a spin analogue of the anharmonic oscillator as a separate working medium, and show that this hitherto unexplored feature of improved performance of refrigeration persists even in this case for both the Otto and Stirling cycles. The improved co-efficient of performance achieved through a higher magnitude of anharmonicity obviously comes at the cost of the energy supplied, as we next show by evaluating the quantitative change in the average energy fluctuation. However, the generic enhancement of performance for the Otto refrigerator grows surprisingly with  increasing energy, thus exhibiting the robustness of anharmonicity as a resource {\it vis-a-vis} the energy cost. Moreover, it can also be seen that the anharmonicity driven improvement of performance for the quantum refrigerator is absent, in general, for the endoreversible classical Otto refrigeration cycle.

{\it Anharmonic Oscillator (AO):-} 
The Hamiltonian for AO with quartic perturbation term can be written as~\citep{Anharmonic,Anharmonic1},
\begin{equation}
H^{ao} = \frac{p^2}{2} + \frac{\omega^2 x^2}{2} + \lambda x^4
\label{Hamiltonian}
\end{equation}
where, $x,p,\omega$ are position, momentum and frequency of the oscillator (see~\citep{Remark}). By choosing units such that the mass of the oscillator, $m=1$, and $\hslash=k_{B}=1$ ($k_B$ being Boltzmann constant), we have,  $x=\frac{a+a^{\dagger}}{\sqrt{2\omega}}$, $p=\frac{a-a^{\dagger}}{i\sqrt{2\omega}}$ in terms of creation ($a^{\dagger}$) and annihilation ($a$) operators and the dimension of $\lambda$ $\sim$ (frequency)$^3$. Thus, the dimensionless variable, $\frac{\lambda}{\omega_0^3}$ ($0 < \frac{\lambda}{\omega_0^3} < 1$) serves as the entity of anharmonicity, where $\omega_0$ is a constant characteristic frequency pertaining to $\lambda$. The energy eigenvalues of the Hamiltonian in Eq.(\ref{Hamiltonian}) correct upto first order in $\lambda$ have the form,
\begin{equation}
E_n=(n+\frac{1}{2})\omega + \frac{3\lambda}{2\omega^2} (n^2 + n + \frac{1}{2})
\label{energy}
\end{equation}
where $n$ is any non-negative integer. Such approximation is useful in terms of accuracy as long as $\lambda$ is quite small. The canonical partition function of AO upto first-order in $\lambda$ turns out to be,
\begin{align}
Z^{ao} = \frac{1}{2} \text{csch}(\frac{\beta \omega}{2}) \Big[ 1 - \frac{3\beta \lambda}{4\omega^2} \coth^2(\frac{\beta \omega}{2})\Big],
\end{align}
where $\beta$ is the inverse temperature of the system.

{\it AO-like spin system:-} The operator form of AO can be simulated in terms of ladder operators of spin angular momentum for spin-$\frac{1}{2}$ particles~\citep{thomas}. In order to make the normal ordered form of Hamiltonian consistent with that of AO, we consider that a spin-$\frac{1}{2}$ particle is placed in a magnetic field $B_z$ along the $z$-direction and  a constant driving Hamiltonian ($\Omega \openone_2$) acts on the system. Hence, the total Hamiltonian of the system is,
\begin{equation}
H^{sp} = \gamma B_z S_z + \Omega \openone_2
\end{equation}
where, the constants $\gamma$ and $\Omega$ can be represented in terms of $\lambda$ and $\omega$ of AO as, $\gamma= \frac{1}{B_z} (\omega+\frac{3\lambda}{\omega^2})$ and $\Omega= (\omega+\frac{9\lambda}{4\omega^2})$. By decomposing the $z$-component of spin angular momentum, $S_z$ (=$\frac{\sigma_z}{2}$) by means of ladder operators $S_+$ and $S_-$, and taking normal order{\footnote{In terms of ($a$, $a^{\dagger}$), the normal ordered form of $x^4$ is given by $(a^{\dagger})^4 + 4 (a^{\dagger})^3 a + 6 (a^{\dagger})^2 + 6 (a^{\dagger})^2 a^2 + 12 a^{\dagger} a + 4 a^{\dagger} a^3 + 6 a^2 + a^4 +3$. We replace $a^{\dagger}$ and $a$ by the spin raising and  lowering operators $S_+ = \begin{pmatrix}
0 & 1\\
0 & 0\\
\end{pmatrix}$, and $S_- = \begin{pmatrix}
0 & 0\\
1 & 0\\
\end{pmatrix}$, respectively, and use $(S_+)^2 = (S_-)^2 =0$.}}, we get
\begin{equation}
H^{sp} = (S_+ S_- + \frac{1}{2}) \omega + (4 S_+ S_- +1) \frac{3\lambda}{4\omega^2}
\label{spin-en}
\end{equation}
which has two energy eigenvalues corresponding to the two levels,  $E_0= \frac{\omega}{2} + \frac{3\lambda}{4\omega^2}$ and $E_1= \frac{3\omega}{2} + \frac{15\lambda}{4\omega^2}$. It is straightforward to obtain the corresponding partition function given by, $Z^{sp}=\sum_{n=0}^1 \exp(-\beta E_n)$. Here too, $\lambda$ is of dimension (frequency)$^3$, and we treat $\frac{\lambda}{\omega_0^3} \in (0,1)$ as the parameter of anharmonicity, with  $\omega_0$ a non-negative constant.

\begin{figure} [H]
\centering
\begin{tabular}{cccc}
\includegraphics[width=0.24\textwidth]{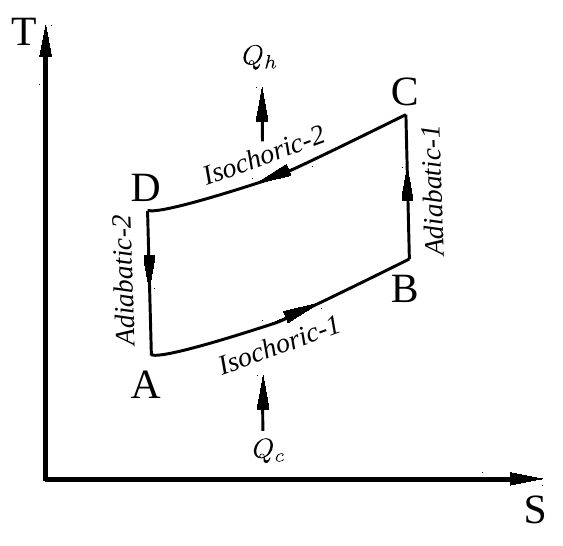} \label{Otto} &
\includegraphics[width=0.24\textwidth]{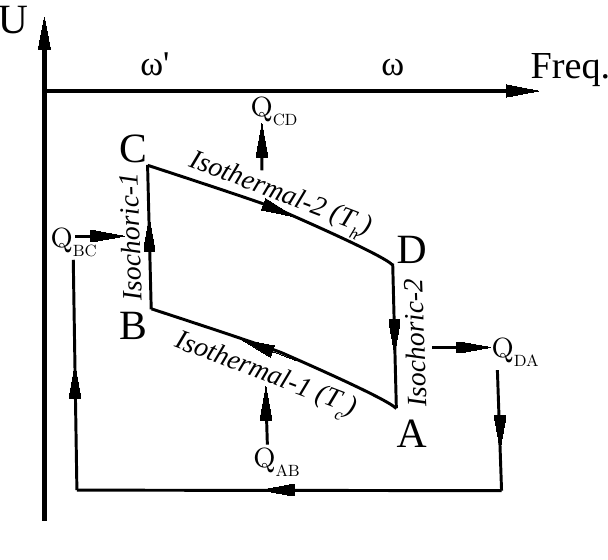} \label{Stirling} \\
\textbf{(a)}  & \textbf{(b)}  \\[6pt]
\end{tabular}
\caption{\footnotesize \textbf{(a)} Temperature-Entropy (T-S) diagram of the 4-step Quantum Otto  cycle. 
\textbf{(b)} Schematic diagram of internal energy(U) vs frequency of the system for the 4-step Quantum Stirling  cycle.}
\label{Otto-Stir}
\end{figure}

{\it Quantum Otto cycle:-}
The four-step Otto refrigerator~\citep{quan1,thomas,thomas1,KR17} (see Fig.\ref{Otto-Stir}) can be described as follows: (i) {\it Isochoric-1 (A$\rightarrow$B):} The system is coupled to a cold reservoir maintained at temperature $T_c$ while the system Hamiltonian is kept fixed at $H'$. The amount of heat absorbed from the cold bath during isochoric cooling is given by~\citep{notes},
\begin{equation}
Q_c = \sum_{n} E_{n}^{'c} (P_{n}^{h} - P_{n}^{c}) > 0,
\end{equation}
where $E_{n}^{'c}=E_{n}|_{\omega=\omega'}$ is either of the form given by Eq.(\ref{energy}) for AO, or of the form of the eigen-energy of the spin system, depending on the case considered. $P_{n}^{c}= \frac{\exp(-\beta E_{n})}{Z^{ao(sp)}}|_{\beta=\beta_c,\omega=\omega'}$ and $P_{n}^{h}= \frac{\exp(-\beta E_{n})}{Z^{ao(sp)}}|_{\beta=\beta_h}$ are the occupation probabilities of the system in the n-th eigenstate corresponding to the points $A$ and $B$, respectively. (ii) {\it Adiabatic-1 (B$\rightarrow$C):} As the process conserves entropy (S) at points B and  C, i.e., $S_B=S_C$, the occupation distribution remains invariant under the adiabatic evolution which alters the Hamiltonian from $H^{ao(sp)}(B)=H'$ to $H^{ao(sp)}(C)=H$ (or frequency  from $\omega'$ to $\omega$) adiabatically. (iii) {\it Isochoric-2 (C$\rightarrow$D):} The system rejects heat to the hot reservoir at temperature $T_{h}$ during isochoric heating, keeping the Hamiltonian fixed at $H$, by an amount~\citep{notes},
\begin{equation}
Q_h = \sum_{n} E_{n}^{h} (P_{n}^{c} - P_{n}^{h}) < 0,
\end{equation}
where $E_{n}^{h}=E_{n}|_{\omega=\omega}$ having the corresponding forms for the AO and spin qubit cases, respectively.  $P_{n}^{h}$ and $P_{n}^{c}$ are the occupation probabilities for the system to remain in the $n$-th eigenstate corresponding to points C and D, respectively. (iv) {\it Adiabatic-2 (D$\rightarrow$A):} During this process the Hamiltonian changes from $H^{ao(sp)}(D)=H$ to $H^{ao(sp)}(A)=H'$ (or frequency  from $\omega$ to $\omega'$) quasi-statically keeping the entropy constant for points D and A, $S_D=S_A$ which, in turn, keeps the occupancies unaltered. The net work done on the system per cycle can be calculated as, $W_O = Q_h + Q_c < 0$ ($|Q_c| < |Q_h|$). The co-efficient of performance (COP) for Otto refrigerator, which is the ratio of heat removed from the cold reservoir ($Q_c$) to the total amount of work ($W_O$) done on the system, is given by
\begin{equation}
\epsilon_O = \frac{Q_c}{|W_O|}
\label{Otto-cop}
\end{equation}

{\it Quantum Stirling cycle:-}
The four-step Stirling refrigerator~\citep{stirling,stirling1} (see Fig.\ref{Otto-Stir}) is as follows: (i) {\it Isothermal-1 (A$\rightarrow$B):} The system is coupled to a cold reservoir maintained at temperature $T_c$, and the Hamiltonian of the system changes slowly from $H^{ao(sp)}(A)=H$ to $H^{ao(sp)}(B)=H'$ (or frequency from $\omega$ to $\omega'$) to keep the system in thermal equilibrium with the cold bath. Meanwhile, the entropy of the system changes from $S_A = S^{ao(sp)}|_{\beta = \beta_c}$ to $S_B = S^{ao(sp)}|_{\beta = \beta_c, \omega=\omega'}$. The amount of heat thereby absorbed is
\begin{equation}
Q_{AB} = T_c (S_B - S_A) > 0,
\label{qAB}
\end{equation}
(ii) {\it Isochoric-1 (B$\rightarrow$C):} The Hamiltonian is kept fixed at $H'$ while the temperature of the system increases from $T_c$ to $T_h$. The mean internal energy of the system increases from $U_B = U^{ao(sp)}|_{\beta = \beta_c, \omega=\omega'}$ to $U_C = U^{ao(sp)}|_{\beta = \beta_h, \omega=\omega'}$. As a result, the system gains heat by an amount~\citep{notes},
\begin{equation}
Q_{BC} = U_C - U_B > 0,
\label{qBC}
\end{equation}
(iii) {\it Isothermal-2 (C$\rightarrow$D):} The system is now attached to a hot reservoir at temperature $T_{h}$ and the quasi-static change in the Hamiltonian from $H^{ao(sp)}(C)=H'$ to $H^{ao(sp)}(D)=H$ (or frequency  from $\omega'$ to $\omega$) governs the change in entropy from $S_C \equiv S^{ao(sp)}|_{\beta = \beta_h, \omega=\omega'}$ to $S_D \equiv S^{ao(sp)}|_{\beta = \beta_h}$. Thus, the heat rejected to the bath is given by,
\begin{equation}
Q_{CD} = T_h (S_D - S_C) < 0,
\label{qCD}
\end{equation}
(iv) {\it Isochoric-2 (D$\rightarrow$A):} The system Hamiltonian remains constant at $H$,  and the temperature changes from $T_h$ to $T_c$ leading to the decrease in mean internal energy from $U_D = U^{ao(sp)}|_{\beta = \beta_h}$ to $U_A = U^{ao(sp)}|_{\beta = \beta_c}$. Thus, the released heat amounts to~\citep{notes}
\begin{equation}
Q_{DA} = U_A - U_D < 0.
\label{qDA}
\end{equation}
In each case the internal energy and entropy can be derived from the partition function as, $U^{ao(sp)} = -\frac{\partial}{\partial \beta} \ln Z^{ao(sp)}$ and $S^{ao(sp)} = \ln Z^{ao(sp)} + \beta U^{ao(sp)}$. The Stirling refrigeration cycle is a regenerative cycle as the input in the isochoric-1 process comes via plugging in the output of the isochoric-2 process. The net work done on the system is $W_S = Q_{AB} + Q_{BC} + Q_{CD} + Q_{DA} < 0$ ($|Q_{AB}| + |Q_{BC}| < |Q_{CD}| + |Q_{DA}|$). The co-efficient of performance (COP) of the Stirling refrigerator is given by,
\begin{equation}
\epsilon_S = \frac{Q_{AB} + Q_{BC}}{|W_S|}
\label{St-cop}
\end{equation}

{\it Improved COP of Otto refrigerator:-}
Ingraining anharmonicity in the oscillator and spin-$\frac{1}{2}$ system, the amount of heat absorbed from the cold bath during the Isochoric-1 process, is respectively\footnote{In the Eqs.(\ref{Qcao}-\ref{Essp}) below the co-efficients of $\lambda$ are given in the {\it Supplemental Material}~\citep{Sup}.}
\begin{align}
Q_c^{ao} =& \frac{\omega'}{2} \Big(\coth\Big[\frac{\beta_h \omega}{2}\Big] - \coth\Big[\frac{\beta_c \omega'}{2}\Big] \Big) + \lambda~ Q_{c1}^{ao}(\omega, \omega', \beta_h, \beta_c) \nonumber\\
&+ O({\lambda}^2), \label{Qcao} \\
Q_c^{sp} =& \frac{\omega'}{2} \Big( \tanh\Big[\frac{\beta_c \omega'}{2} \Big] - \tanh\Big[\frac{\beta_h \omega}{2} \Big] \Big) + \lambda~ Q_{c1}^{sp}(\omega, \omega', \beta_h, \beta_c) \nonumber\\
&+ O({\lambda}^2). \label{Qcsp}
\end{align}
The heat rejected to the hot reservoir during the Isochoric-2 process by the AO and qubit are respectively,
\begin{align}
Q_h^{ao} =& -\frac{\omega}{2} \Big(\coth\Big[\frac{\beta_h \omega}{2}\Big] - \coth\Big[\frac{\beta_c \omega'}{2}\Big] \Big) +\lambda~ Q_{h1}^{ao}(\omega, \omega', \beta_h, \beta_c) \nonumber\\
&+ O({\lambda}^2), \label{Qhao} \\
Q_h^{sp} =& -\frac{\omega}{2} \Big( \tanh\Big[\frac{\beta_c \omega'}{2} \Big] - \tanh\Big[\frac{\beta_h \omega}{2} \Big]\Big) +\lambda~ Q_{h1}^{sp}(\omega, \omega', \beta_h, \beta_c) \nonumber\\
&+ O({\lambda}^2). \label{Qhsp}
\end{align}
Therefore,  the COPs of the Otto refrigerator corresponding to the AO and spin system respectively become,
\begin{align}
\epsilon_O^{ao} =& \frac{\omega'}{\omega-\omega'} + \frac{3 \lambda}{2\omega^2 \omega'^2} \frac{(\omega^3 - \omega'^3)}{(\omega-\omega')^2} \Big( \coth\Big[\frac{\beta_h \omega}{2}\Big] + \coth\Big[\frac{\beta_c \omega'}{2}\Big]\Big) \nonumber\\
&+ O({\lambda}^2), \label{Eoao} \\
\epsilon_O^{sp} =& \frac{\omega'}{\omega-\omega'} + \frac{3\lambda}{2\omega^2 \omega'^2} \frac{(\omega^3 - \omega'^3)}{(\omega-\omega')^2} + O({\lambda}^2). \label{Eosp}
\end{align}

The negative work condition of the Otto refrigerator for both AO and spin systems dictates, $\omega > \omega'$ and $\beta_c \omega' > \beta_h \omega$, making the co-efficients of $\lambda$ ($\lambda > 0$) in the expressions of $\epsilon_{O}^{ao}$ and $\epsilon_{O}^{sp}$ positive (see, {\it Supplemental Material}~\citep{Sup} for further details). Hence, the COP of the Otto refrigerator corresponding to both the working substances monotonically increases with the anharmonicity parameter (for the harmonic oscillator it is essentially $\frac{\omega'}{\omega-\omega'}$). It may be further noted from the functional forms in Eqs.(\ref{Eoao}-\ref{Eosp})  that the COP for AO is higher than the COP for the spin system for all $\lambda$. 

{\it Improved COP of Stirling refrigerator:-}
Corresponding to the change in frequency ($\omega\rightarrow\omega'$) during the Isothermal-1 process at inverse temperature, $\beta_c$, the absorbed heat from the cold bath by AO and qubit are respectively,
\begin{align}
Q_{AB}^{ao} =& \frac{\omega'}{2} \coth[\frac{\beta_c \omega'}{2}] - \frac{\omega}{2} \coth[\frac{\beta_c \omega}{2}] + \frac{1}{\beta_c} \ln \Big[ \frac{\sinh[\frac{\beta_c \omega}{2}]}{\sinh[\frac{\beta_c \omega'}{2}]} \Big]\nonumber\\
& + \lambda~ Q_{AB1}^{ao}(\omega, \omega', \beta_h, \beta_c) + O({\lambda}^2), \label{Qabao}\\
Q_{AB}^{sp} =& \frac{\omega}{2} \tanh[\frac{\beta_c \omega}{2}] - \frac{\omega'}{2} \tanh[\frac{\beta_c \omega'}{2}] + \frac{1}{\beta_c} \ln \Big[ \frac{\cosh[\frac{\beta_c \omega'}{2}]}{\cosh[\frac{\beta_c \omega}{2}]} \Big] \nonumber\\
& + \lambda~ Q_{AB1}^{sp}(\omega, \omega', \beta_h, \beta_c) + O({\lambda}^2). \label{Qabsp}
\end{align}
The heat absorbed by AO and qubit during Isochoric-1 process due to rise in temperature turns out to be,
\begin{align}
Q_{BC}^{ao} =& \frac{\omega'}{2} \Big(\coth[\frac{\beta_h \omega'}{2}] - \coth[\frac{\beta_c \omega'}{2}] \Big)+ \lambda\; Q_{BC1}^{ao}(\omega, \omega', \beta_h, \beta_c)\nonumber\\
&+ O({\lambda}^2), \label{Qbcao} \\
Q_{BC}^{sp} =& \frac{\omega'}{2} \Big( \tanh[\frac{\beta_c \omega'}{2}] - \tanh[\frac{\beta_h \omega'}{2}] \Big) + \lambda\; Q_{BC1}^{sp}(\omega, \omega', \beta_h, \beta_c)\nonumber\\
&+ O({\lambda}^2). \label{Qbcsp}
\end{align}
The heat rejected to the bath during Isothermal-2 process at inverse temperature, $\beta_h$ by AO and spin system are respectively,
\begin{align}
Q_{CD}^{ao} =& -\frac{\omega'}{2} \coth[\frac{\beta_h \omega'}{2}] + \frac{\omega}{2} \coth[\frac{\beta_h \omega}{2}] - \frac{1}{\beta_h} \ln \Big[ \frac{\sinh[\frac{\beta_h \omega}{2}]}{\sinh[\frac{\beta_h \omega'}{2}]} \Big] \nonumber\\
&+ \lambda~ Q_{CD1}^{ao}(\omega, \omega', \beta_h, \beta_c) + O({\lambda}^2), \label{Qcdao} \\
Q_{CD}^{sp} =& -\frac{\omega}{2} \tanh[\frac{\beta_h \omega}{2}] + \frac{\omega'}{2} \tanh[\frac{\beta_h \omega'}{2}] - \frac{1}{\beta_h} \ln \Big[ \frac{\cosh[\frac{\beta_h \omega'}{2}]}{\cosh[\frac{\beta_h \omega}{2}]} \Big] \nonumber\\
&+ \lambda~ Q_{CD1}^{sp}(\omega, \omega', \beta_h, \beta_c) + O({\lambda}^2). \label{Qcdsp}
\end{align}
Finally, in the Isochoric-2 process, the heat rejected by AO and spin are respectively,
\begin{align}
Q_{DA}^{ao} &= -\frac{\omega}{2} \Big(\coth[\frac{\beta_h \omega}{2}] - \coth[\frac{\beta_c \omega}{2}] \Big)+  \lambda\; Q_{DA1}^{ao}(\omega, \omega', \beta_h, \beta_c)\nonumber\\
&+ O({\lambda}^2), \label{Qdaao} \\
Q_{DA}^{sp} &= -\frac{\omega}{2} \Big(\tanh[\frac{\beta_c \omega}{2}] - \tanh[\frac{\beta_h \omega}{2}] \Big) +  \lambda\; Q_{DA1}^{sp}(\omega, \omega', \beta_h, \beta_c)\nonumber\\
&+ O({\lambda}^2). \label{Qdasp}
\end{align}
From the Eqs.(\ref{Qabao}-\ref{Qdasp}), we obtain the COP of Stirling refrigerator for AO and qubit system respectively as,
\begin{align}
 \epsilon_S^{ao} &= \frac{\omega'~ \coth[\frac{\beta_h \omega'}{2}] - \omega~ \coth[\frac{\beta_c \omega}{2}] + \frac{1}{\beta_c} \ln \Big[ \frac{\sinh^2[\frac{\beta_c \omega}{2}]}{\sinh^2[\frac{\beta_c \omega'}{2}]} \Big]}{\frac{1}{\beta_h} \ln \Big[ \frac{\sinh^2[\frac{\beta_h \omega}{2}]}{\sinh^2[\frac{\beta_h \omega'}{2}]} \Big] + \frac{1}{\beta_c} \ln\Big[ \frac{\sinh^2[\frac{\beta_c \omega'}{2}]}{\sinh^2[\frac{\beta_c \omega}{2}]} \Big]} \nonumber\\
& + \lambda \;\epsilon_{S1}^{ao}(\omega, \omega', \beta_h, \beta_c) + O({\lambda}^2), \label{Esao}\\
 \epsilon_S^{sp} =& \frac{\omega~ \tanh[\frac{\beta_c \omega}{2}] - \omega'~ \tanh[\frac{\beta_h \omega'}{2}] + \frac{1}{\beta_c} \ln \Big[ \frac{\cosh^2[\frac{\beta_c \omega'}{2}]}{\cosh^2[\frac{\beta_c \omega}{2}]} \Big]}{\frac{1}{\beta_h} \ln \Big[ \frac{\cosh^2[\frac{\beta_h \omega'}{2}]}{\cosh^2[\frac{\beta_h \omega}{2}]} \Big] + \frac{1}{\beta_c} \ln\Big[ \frac{\cosh^2[\frac{\beta_c \omega}{2}]}{\cosh^2[\frac{\beta_c \omega'}{2}]} \Big]} \nonumber\\
& + \lambda \;\epsilon_{S1}^{sp}(\omega, \omega', \beta_h, \beta_c) + O({\lambda}^2). \label{Essp}
\end{align}

The negative work condition for the Stirling refrigerator,  $\omega > \omega'$, $\beta_h < \beta_c$ makes the functions $\epsilon_{S1}^{ao(sp)}$ positive in Eq.(\ref{Esao}-\ref{Essp}) (see, {\it Supplemental Material}~\citep{Sup} for further details). The COP of the Stirling refrigerator thereby increases w.r.t. the dimensionless anharmonicity parameter $0 \leq \frac{\lambda}{\omega_0^3} \leq 1$ for both the working substances (anharmonic oscillator as well as the analogous spin-$\frac{1}{2}$ system) compared to that of the harmonic oscillator.

{\it Energy cost for improved refrigeration:-} The physical reason for the increase in COP with $\lambda$ is the change in the Hamiltonian which can be driven externally. So, it is imperative to analyse the performance of refrigerators due to average fluctuation in the energy at thermal equilibrium responsible for anharmonicity of a working medium. This average calculated over all possible energy eigenstates, is called {\it "energy cost"} in the context of thermodynamic cycles~\citep{nonlin3}. For the anharmonic oscillator, it is given by
\begin{align}
\delta H & := \sum_{n=0}^{\infty} \frac{\exp(-\beta E_n)}{Z^{ao}} [E_n - (n+\frac{1}{2}) \omega] \nonumber\\
&= \frac{1}{Z^{ao}} \frac{3\lambda}{8 \omega^2} \text{csch}(\frac{\beta \omega}{2}) \coth^2(\frac{\beta \omega}{2}) + O({\lambda}^2),
\end{align}
and for the AO-like spin system,
\begin{align}
\delta H & := \sum_{n=0}^{1} \frac{\exp(-\beta E_n)}{Z^{sp}} [E_n - (n+\frac{1}{2}) \omega] \nonumber\\
&= \frac{1}{Z^{sp}} \frac{3\lambda}{4\omega^2} \exp(-\frac{3\beta \omega}{2})\; (5+\exp(\beta \omega)) + O({\lambda}^2).
\end{align}

\begin{figure} [H]
\centering
\begin{tabular}{cccc}
\includegraphics[width=0.24\textwidth]{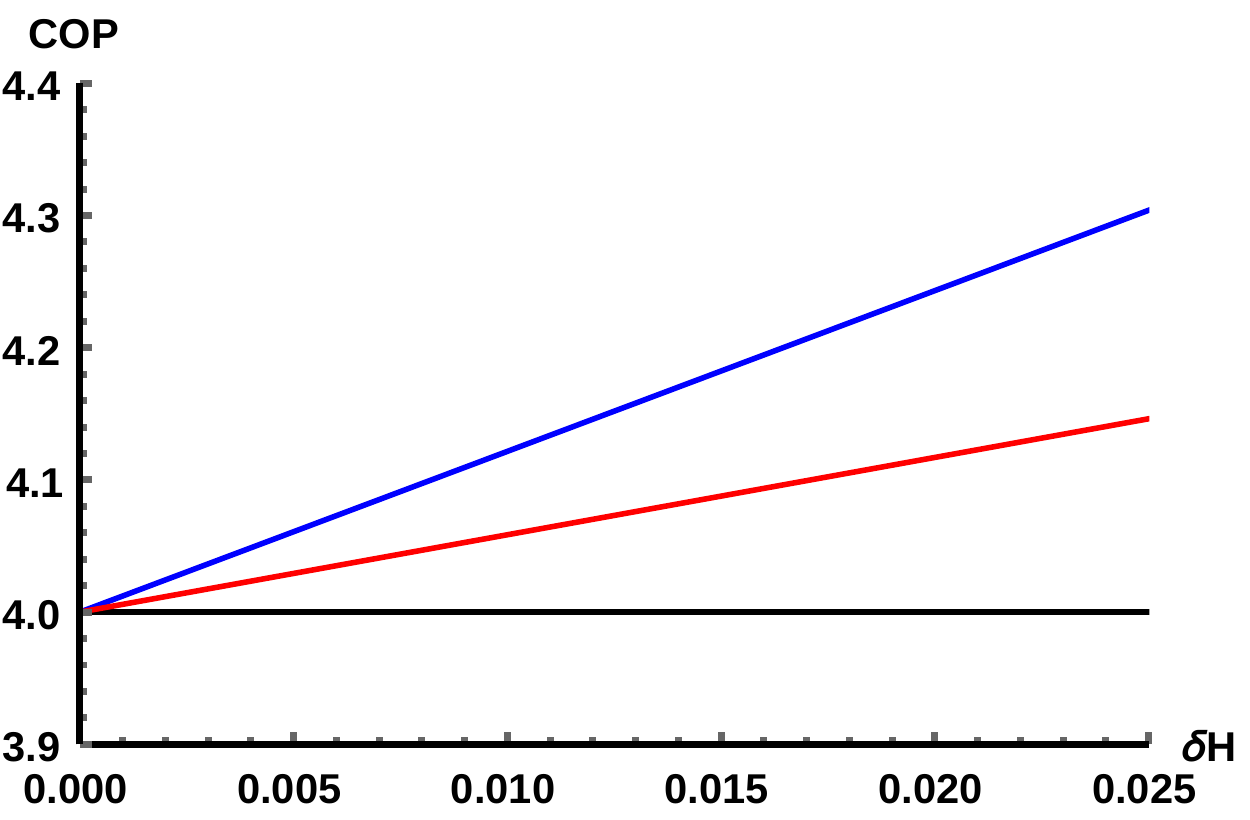} &
\includegraphics[width=0.24\textwidth]{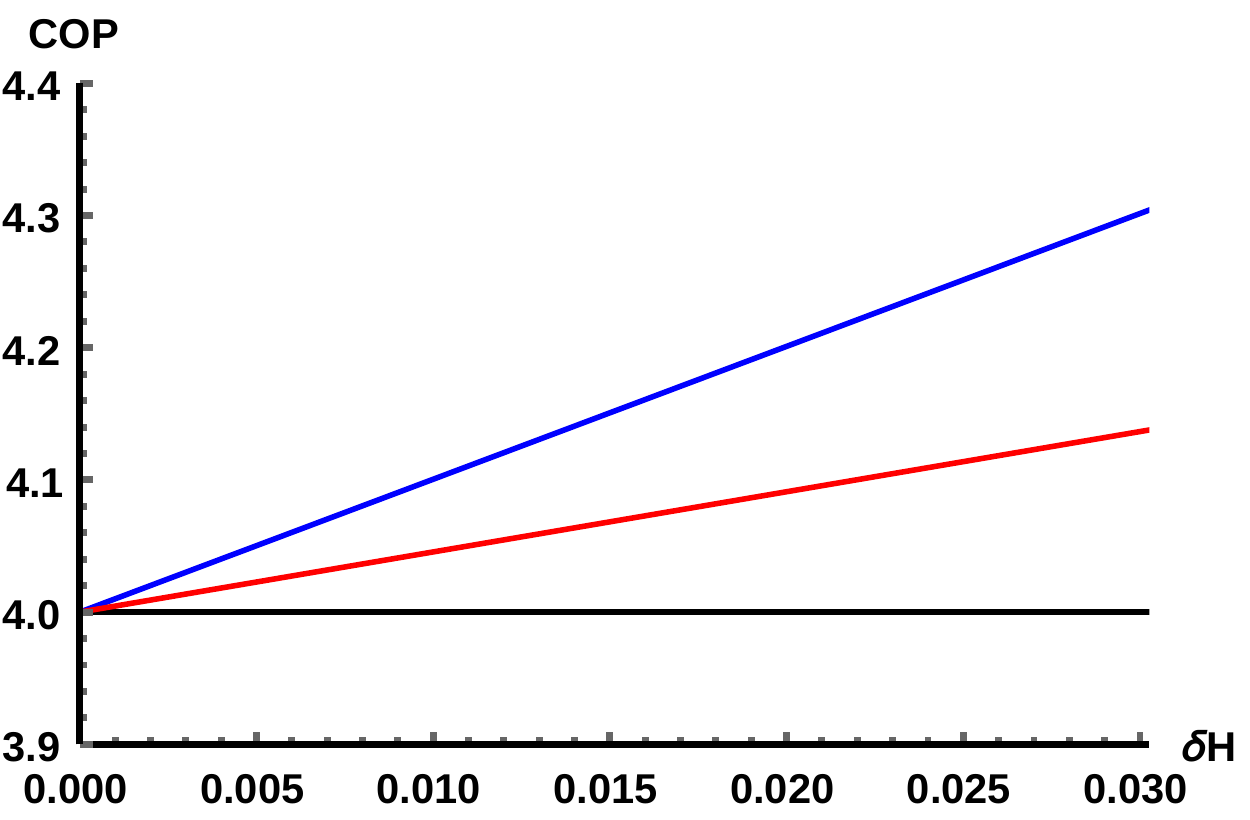} \\
\textbf{(a)}  & \textbf{(b)}  \\[6pt]
\end{tabular}
\caption{\footnotesize (Color Online)  \textbf{(a)} and \textbf{(b)} indicate COP of the Otto refrigerator($\epsilon_O$) as a function of $\delta H$ using case (i) and case (ii), respectively. The upper (blue), middle (red), lower (black) lines imply the working media as AO, AO-like qubit and harmonic oscillator, respectively. The  parameters are $\beta_h= \frac{1}{2}$, $\beta_c =1$, $\omega$=5, $\omega'$=4 and $\omega_0 =\sqrt[3]{0.6}$.}
\label{ec1}
\end{figure}

\begin{figure} [H]
\centering
\begin{tabular}{cccc}
\includegraphics[width=0.24\textwidth]{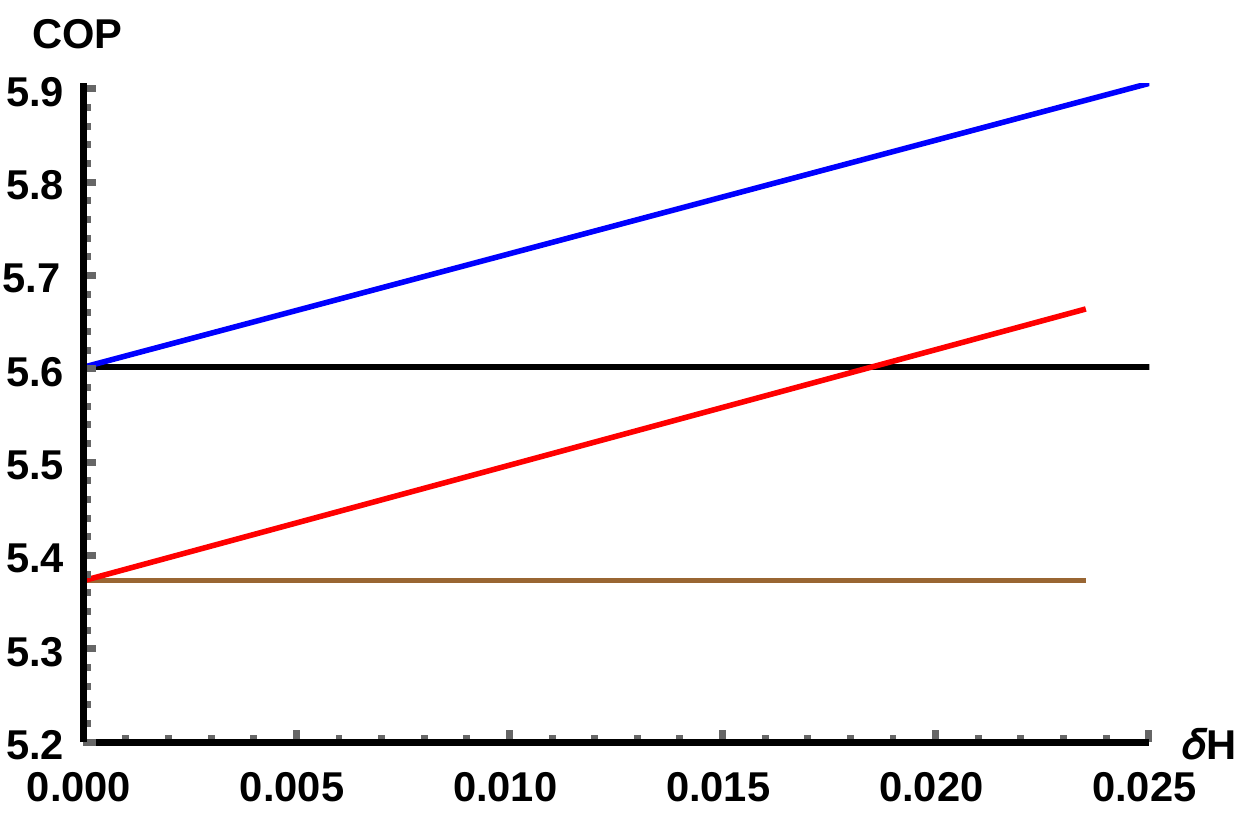} &
\includegraphics[width=0.24\textwidth]{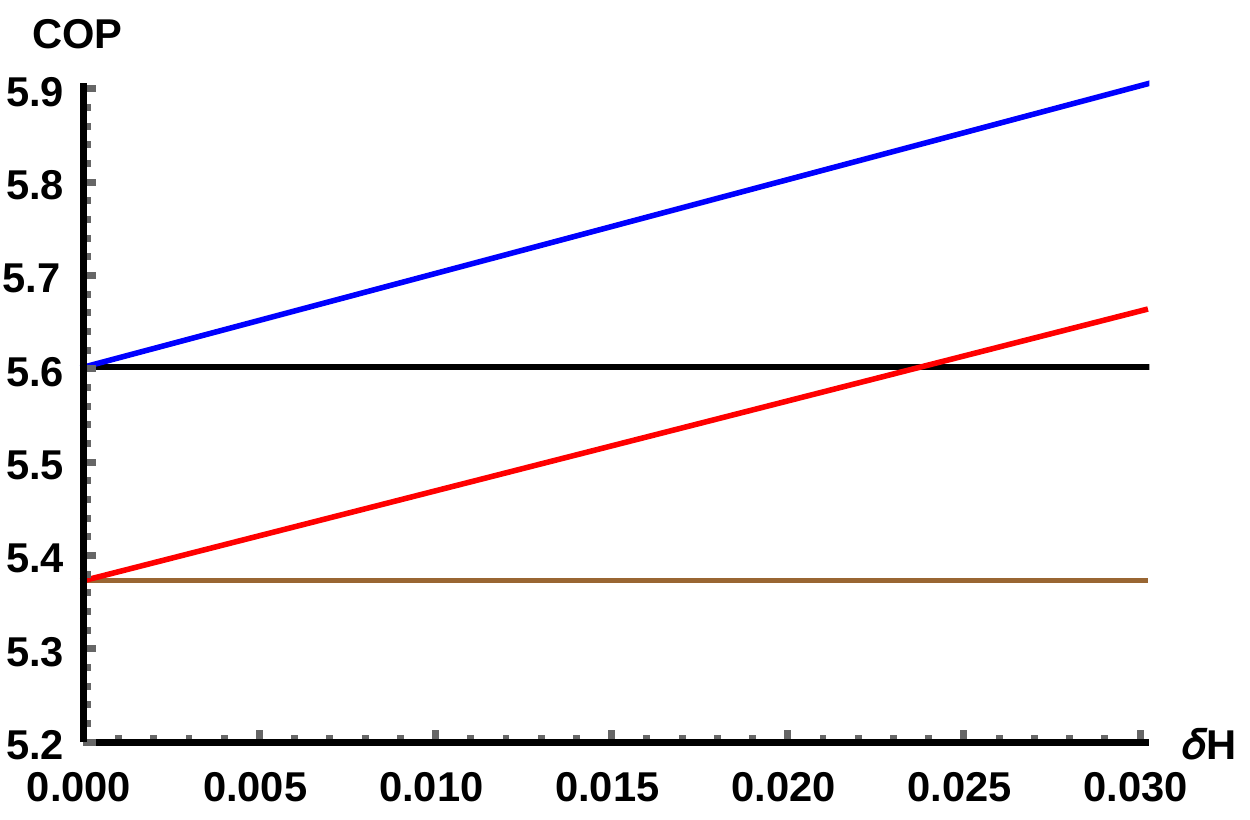} \\
\textbf{(a)}  & \textbf{(b)}  \\[6pt]
\end{tabular}
\caption{\footnotesize (Color Online) COP ($\epsilon_S$) vs energy cost for the Stirling refrigerator, \textbf{(a)} corresponding to case (i), and \textbf{(b)} corresponding to case (ii). The upper (blue) and lower (red) curves, and the upper (black) and lower (brown) straight  lines indicate the working media for AO, AO-like qubit system, harmonic oscillator and qubit analogous to harmonic oscillator, respectively. The  parameters are $\beta_h= \frac{1}{2}$, $\beta_c =1$, $\omega$=5, $\omega'$=4 and $\omega_0 =\sqrt[3]{0.6}$.}
\label{ec2}
\end{figure}

We study two cases, namely (i) $\beta = \beta_h$, and (ii) $\beta = \beta_c$, $\omega = \omega'$, to observe the variation of COP compared to $\delta H$. For the choice of parameters: $\beta_h = \frac{1}{2}$, $\beta_c = 1$, $\omega = 5$, $\omega' = 4$ and $\omega_0 = \sqrt[3]{0.6}$, the variation of $\delta H$ corresponding to the range $\frac{\lambda}{\omega_0^3} \in (0,1)$ lies in the interval $[0,0.025]$ for the AO, and $[0,0.023]$ for the qubit, for case (i), and in  $[0,0.030]$ for both AO and qubit for case (ii). The COP for Otto and Stirling refrigerators are plotted against $\delta H$ in Fig.\ref{ec1} and Fig.\ref{ec2} respectively, corresponding  to the case (i) and case (ii) above. In all the cases, COP increases with $\delta H$ compared to the harmonic counterparts (with $\delta H$=0 (or $\lambda$=0)). The improvement of COP is higher for AO compared to the AO-like qubit. Thus, from both Fig.\ref{ec1} as well as Fig.\ref{ec2}, it is clear that for small values of the anharmonicity parameter $\lambda$, anharmonic oscillator- based (together with its spin counterpart) Otto as well as Stirling refrigerators are energetically more efficient -- so far as the respective COPs of the refrigerators are concerned -- compared to the the case where there is no anharmonicity. The vailidity of our results extends upto a threshold value of $\lambda$ -- the value upto which the energy spectrum in Eq.(\ref{energy}) of the anharmonic oscillator Hamiltmonian in Eq.(\ref{Hamiltonian}) is valid.

{\it Discussions:-}
To summarize, in this work we have shown that for small quartic perturbation, anharmonic substances outperform harmonic ones in terms of the COP ($\epsilon$) for both the Otto and Stirling quantum refrigerators. The improvement in COP persists with increase of the energy cost to create anharmonicity for the oscillator as well as the analogous spin qubit system. This establishes anharmonicity as a resource for certain thermodynamic processes at the quantum level. Our results, thus add another significant element to some physical  perspectives discussed in the context of quantum thermodynamical resource theories~\citep{r1,r2}. 

It may be noted here that COP of the endoreversible Otto refrigerator~\citep{endo} modelled using the classical harmonic oscillator is the same as that for the quantum harmonic oscillator~\citep{Deffner}. However, it can be shown that anharmonicity in the classical oscillator  may lead to diminishing of COP (see, {\it Supplemental Material}~\citep{Sup} for further details). As the Carnot limit of COP for the classical refrigerator is lower than the COP of quantum Otto refrigerator – irrespective of quantum harmonic oscillator or corresponding spin-$\frac{1}{2}$ system as the medium --  hence, anharmonicity at the quantum level provides further improvement. A possible contributing factor for such enhancement in COP could be the different ratio of change in the energy gaps for the adiabatic processes using anharmonic media \citep{Egap} compared to the harmonic media which maintain the same ratio between the energy gaps \citep{quan1}, i.e. $\frac{\omega}{\omega'}$.

In order to compare the refrigerator performance {\it vis-a-vis} the energy cost for the anharmonic oscillator with the spin-$\frac{1}{2}$ particle, the slopes of $\epsilon$ vs $\lambda$ plots may be compared. In case of the Otto refrigerator, $\frac{d \epsilon_O^{ao}}{d\lambda} : \frac{d \epsilon_O^{sp}}{d\lambda} = \coth(\frac{\beta_h \omega}{2}) + \coth(\frac{\beta_c \omega'}{2}) > 1$ for all $\lambda, \omega > \omega'$ and $\beta_h < \beta_c$. This implies that the anharmonic oscillator is more useful than the qubit in terms of growth in COP versus the external driving energy in case of the Otto cycle. However, for the Stirling refrigerator the ratio $\frac{d \epsilon_S^{ao}}{d\lambda} : \frac{d \epsilon_S^{sp}}{d\lambda}$ changes depending on the parameter values $\lambda, \omega, \omega', \beta_h, \beta_c$.

Before concluding, note that our present results should motivate further investigations on modelling particular experimentally driven forms of nonlinearity~\citep{Exp1,Exp2,Exp3,Exp4,Exp5,Exp6,Exp7,Exp8,TI2,OM2} in order to demonstrate practically more enhanced improvement of performance {\it vis-a-vis} the associated energy costs for quantum refrigerators. This may involve more rigorous analysis of higher order perturbations of the oscillator potential~\citep{nonlin3,anharm1,WKB}. Additionally, the domain of superconducting transmonic qubits~\citep{trans} which can be designed by sufficiently enhancing the amount of anharmonicity, should form an avenue for testing our anharmonicity-induced analogous spin-$\frac{1}{2}$ model. Finally, it may be fascinating to explore such effects of nonlinearity in other applicable arenas of active present interest such as in coupled working media~\citep{thomas,thomas1,c1,c2}, or non-Markovian reservoirs~\citep{thomas2,MBM20}. Progress in the above directions may act as a pedestal for establishing a resource theoretic framework for anharmonicity in quantum thermodynamics~\citep{r1,r2,cost}, inspired by other quantum resource theories~\citep{CG19}.

{\it Acknowledgements:}
SD acknowledges financial support through INSPIRE Fellowship from Department of Science and Technology (DST), Govt. of India (Grant No.C/5576/IFD/2015-16). SG thankfully acknowledges the hospitality of S. N. Bose National Centre for Basic Sciences for his visit during which part of the work was done. SG would like to thank George Thomas and Subhashish Banerjee for useful discussions. SG and ASM acknowledge the support from Interdisciplinary Cyber Physical Systems (ICPS) program of  DST, India, Grant No. DST/ICPS/QuEST/Theme-1/2019/13.

\bibliography{KDGM2}

\begin{thebibliography}{65}
\bibitem [{Rem()}]{Remark}%
  \BibitemOpen
  \href@noop {} {\ }\bibinfo {note} {One can, of course, consider the
  non-linear term as $\lambda x^{2n}$ ($n \geq 2$) instead of $\lambda x^{4}$
  given in Eq.(\ref{Hamiltonian}) and get solution similar to
  Eq.(\ref{energy}). In such case, higher value of n demands lower value of
  $\lambda$.}\BibitemShut {Stop}%
\bibitem [{not()}]{notes}%
  \BibitemOpen
  \href@noop {} {\ }\bibinfo {note} {We assume the formalism of weak coupling
  as envisaged in \cite{Alicki} for calculating heat and work}\BibitemShut
  {NoStop}%
\bibitem [{Ega()}]{Egap}%
  \BibitemOpen
  \href@noop {} {\ }\bibinfo {note} {During adiabatic processes, the energy
  gaps change by the ratio, $\mu = \frac{E_n^h - E_m^h}{E_n^{'c} - E_m^{'c}}$
  ($n \neq m$ and $n,m = \lbrace 0,1,2,... \rbrace$). The energy eigenvalues of
  harmonic oscillator, i.e. $E_n = (n+\frac{1}{2}) \omega$ manifests that, $\mu
  = \frac{\omega}{\omega'} ~\forall n,m$. Whereas using Eq.(\ref{energy}) for
  anharmonic oscillator, we find that, $\mu = \frac{\omega + (n+m+1)
  \frac{3\lambda}{2 \omega^2}}{\omega' + (n+m+1) \frac{3\lambda}{2 \omega'^2}}$
  which is not equal for every energy gap. For example, ($n=0,m=1$) implies
  $\mu =
  \frac{\omega+\frac{3\lambda}{\omega^2}}{\omega'+\frac{3\lambda}{\omega'^2}}$
  while ($n=1,m=2$) implies $\mu =
  \frac{\omega+\frac{6\lambda}{\omega^2}}{\omega'+\frac{6\lambda}{\omega'^2}}$.}\BibitemShut
  {Stop}%
\end{thebibliography}

\begin{widetext}
\section{Supplemental Material}
\beginsupplement

The supplemental material is organised as follows: In Sec.\ref{A1}, the explicit forms of the co-efficients of $\lambda$ in the expressions from Eq.(\ref{Qcao}) to Eq.(\ref{Essp}) of the main text are provided. In Sec.\ref{A2}, the Co-efficient of Performance (COP) for the anharmonic oscillator as working medium corresponding to the classical Otto refrigerator is compared to that of the quantum Otto refrigerator.

\section{Expressions for the co-efficients of $\lambda$ given in Eq.(\ref{Qcao})-(\ref{Essp})} \label{A1}

Co-efficient of $\lambda$ in Eq.(14):
\begin{align}
Q_{c1}^{ao}(\omega,\omega',\beta_h,\beta_c) = \frac{3}{4 \omega^2 \omega'^2} \Bigg(\omega^2 (\beta_c \omega' \coth (\frac{\beta_c \omega'}{2})-1) \text{csch}^2(\frac{\beta_c \omega'}{2})+\text{csch}^2(\frac{\omega \beta_h}{2}) (\omega^2-\beta_h (\omega')^3 \coth (\frac{\omega \beta_h}{2}))\Bigg)
\end{align}

Co-efficient of $\lambda$ in Eq.(15):
\begin{align}
Q_{c1}^{sp}(\omega,\omega',\beta_h,\beta_c) = \frac{3}{4 \omega^2 \omega'^2} \Bigg(\omega^2 (\tanh (\frac{\beta_c \omega'}{2})+\beta_c \omega' \text{sech}^2(\frac{\beta_c \omega'}{2})-\tanh (\frac{\omega \beta_h}{2}))-\beta_h \omega'^3 \text{sech}^2 (\frac{\omega \beta_h}{2})\Bigg)
\end{align}

Co-efficient of $\lambda$ in Eq.(16):
\begin{align}
Q_{h1}^{ao}(\omega,\omega',\beta_h,\beta_c) = - \frac{3}{4 \omega ^2 \omega'^2} \Bigg( (\omega^3 \beta_c \coth (\frac{\beta_c \omega'}{2})-\omega'^2) \text{csch}^2(\frac{\beta_c \omega'}{2})- \omega'^2 (\omega \beta_h \coth (\frac{\omega \beta_h}{2})-1) \text{csch}^2(\frac{\omega \beta_h}{2})\Bigg)
\end{align}

Co-efficient of $\lambda$ in Eq.(17):
\begin{align}
Q_{h1}^{sp}(\omega,\omega',\beta_h,\beta_c) = - \frac{3}{4 \omega^2 \omega'^2} \Bigg(\omega^3 \beta_c \text{sech}^2 (\frac{\beta_c \omega'}{2})-\omega'^2 (-\tanh (\frac{\beta_c \omega'}{2})+\tanh (\frac{\omega \beta_h}{2})+\omega \beta_h \text{sech}^2(\frac{\omega \beta_h}{2}))\Bigg)
\end{align}

Co-efficient of $\lambda$ in Eq.(20):
\begin{align}
Q_{AB1}^{ao}(\omega,\omega',\beta_h,\beta_c) = \frac{3}{8} \beta_c \Bigg(\frac{\sinh (\omega  \beta_c) \text{csch}^4(\frac{\omega  \beta_c}{2})}{\omega}-\frac{\sinh (\beta_c \omega') \text{csch}^4(\frac{\beta_c \omega'}{2})}{\omega'}\Bigg)
\end{align}

Co-efficient of $\lambda$ in Eq.(21):
\begin{align}
Q_{AB1}^{sp}(\omega,\omega',\beta_h,\beta_c) = \frac{3 \beta_c}{2 \omega \omega'} \Bigg(\frac{\omega'}{\cosh (\omega \beta_c)+1}-\frac{\omega}{\cosh (\beta_c \omega')+1}\Bigg)
\end{align}

Co-efficient of $\lambda$ in Eq.(22):
\begin{align}
Q_{BC1}^{ao}(\omega,\omega',\beta_h,\beta_c) = \frac{3}{16 \omega'^2} \Bigg(\sinh (\beta_h \omega') (\sinh (\beta_h \omega')-2 \beta_h \omega') \text{csch}^4(\frac{\beta_h \omega'}{2})-\sinh (\beta_c \omega') (\sinh (\beta_c \omega')-2 \beta_c \omega') \text{csch}^4(\frac{\beta_c \omega'}{2})\Bigg)
\end{align}

Co-efficient of $\lambda$ in Eq.(23):
\begin{align}
& Q_{BC1}^{sp}(\omega,\omega',\beta_h,\beta_c) = \frac{3 \text{sech}^2(\frac{\beta_c \omega'}{2}) \text{sech}^2(\frac{\beta_h \omega'}{2})}{8 \omega'^2} \Bigg(\beta_h \omega' \cosh (\beta_c \omega')+\beta_h \omega'+\sinh (\beta_h \omega') -\beta_c \omega'-\sinh (\beta_c \omega') \nonumber\\
& -\sinh (\omega' (\beta_c-\beta_h)) -\beta_c \omega' \cosh (\beta_h \omega')\Bigg)
\end{align}

Co-efficient of $\lambda$ in Eq.(24):
\begin{align}
Q_{CD1}^{ao}(\omega,\omega',\beta_h,\beta_c) = -\frac{3}{8} \beta_h \Bigg(\frac{\sinh (\omega \beta_h) \text{csch}^4(\frac{\omega \beta_h}{2})}{\omega} - \frac{\sinh (\beta_h \omega') \text{csch}^4(\frac{\beta_h \omega'}{2})}{\omega'}\Bigg)
\end{align}

Co-efficient of $\lambda$ in Eq.(25):
\begin{align}
Q_{CD1}^{sp}(\omega,\omega',\beta_h,\beta_c) = -\frac{3 \beta_h}{2 \omega \omega'} \Bigg(\frac{\omega'}{\cosh (\omega \beta_h)+1}-\frac{\omega }{\cosh (\beta_h \omega')+1}\Bigg)
\end{align}

Co-efficient of $\lambda$ in Eq.(26):
\begin{align}
Q_{DA1}^{ao}(\omega,\omega',\beta_h,\beta_c) = -\frac{3}{16 \omega^2} \Bigg(\sinh (\omega \beta_h) (\sinh (\omega \beta_h)-2 \omega  \beta_h) \text{csch}^4(\frac{\omega \beta_h}{2})-\sinh (\omega \beta_c) (\sinh (\omega \beta_c)-2 \omega \beta_c) \text{csch}^4(\frac{\omega \beta_c}{2})\Bigg)
\end{align}

Co-efficient of $\lambda$ in Eq.(27):
\begin{align}
& Q_{DA1}^{sp}(\omega,\omega',\beta_h,\beta_c) = -\frac{3 \text{sech}^2(\frac{\omega \beta_c}{2}) \text{sech}^2(\frac{\omega \beta_h}{2})}{8 \omega^2} \Bigg(\sinh (\omega \beta_c)+\sinh (\omega (\beta_c-\beta_h))+\omega \beta_c (\cosh (\omega \beta_h)+1) \nonumber\\
& -\omega  \beta _h (\cosh (\omega \beta_c)+1)-\sinh (\omega \beta_h)\Bigg)
\end{align}

Co-efficient of $\lambda$ in Eq.(28):
\begin{align}
& \epsilon_{S1}^{ao}(\omega,\omega',\beta_h,\beta_c) = \frac{3 \beta _c \beta _h}{8 \omega ^2 \omega'^2 \Bigg(\beta_c \ln \big[\frac{\text{csch}(\frac{\beta_h \omega'}{2})}{\text{csch}(\frac{\omega \beta_h}{2})}\big] + \beta_h \ln \big[\frac{\text{csch}(\frac{\omega \beta_c}{2})}{\text{csch}(\frac{\beta_c \omega'}{2})}\big]\Bigg)^2} \Bigg( (-\omega \beta_c \beta_h \omega'^2 \coth^3(\frac{\omega \beta_c}{2})+\beta_h \omega'^2 \coth ^2(\frac{\omega \beta_c}{2}) (2 \ln (\text{csch}(\frac{\beta_c \omega'}{2})) \nonumber\\
& -2 \ln (\text{csch}(\frac{\omega \beta_c}{2}))+\beta_c \omega' \coth (\frac{\beta_h \omega'}{2}))-\beta_h \omega'^2 \coth^2(\frac{\omega \beta_h}{2}) (2 \ln (\text{csch}(\frac{\beta_c \omega'}{2}))-2 \ln (\text{csch}(\frac{\omega \beta_c}{2}))+\beta_c \omega' \coth (\frac{\beta_h \omega'}{2})) \nonumber\\
& -\omega (-\omega \beta_c \beta_h \omega' \coth^3(\frac{\beta_h \omega'}{2})+2 \omega \beta_c \coth^2(\frac{\beta_h \omega'}{2}) (\ln (\text{csch}(\frac{\omega \beta_h}{2}))-\ln (\text{csch}(\frac{\beta_h \omega'}{2})))+\omega \beta_c \coth^2(\frac{\beta_c \omega'}{2}) (\beta_h \omega' \coth (\frac{\beta_h \omega'}{2}) \nonumber\\
& +2 \ln (\text{csch}(\frac{\beta_h \omega'}{2}))-2 \ln (\text{csch}(\frac{\omega \beta_h}{2})))+\omega' (\beta_c \ln (\text{csch}(\frac{\beta_h \omega'}{2}))-\beta_h \ln (\text{csch}(\frac{\beta_c \omega'}{2}))-\beta_c \ln (\text{csch}(\frac{\omega \beta_h}{2}))+\beta_h \ln (\text{csch}(\frac{\omega \beta_c}{2}))) \nonumber\\
& (\omega \beta_h \sinh (\beta_h \omega') \text{csch}^4(\frac{\beta_h \omega'}{2})-\beta_c \omega' \sinh (\omega \beta_c) \text{csch}^4(\frac{\omega \beta_c}{2})))+\omega  \beta_c \beta_h \coth (\frac{\omega \beta_c}{2}) (\omega^2 (\coth^2(\frac{\beta_c \omega'}{2})-\coth^2(\frac{\beta_h \omega'}{2})) \nonumber\\
& +\omega'^2 \coth^2(\frac{\omega \beta_h}{2}))) \Bigg)
\end{align}

Co-efficient of $\lambda$ in Eq.(29):
\begin{align}
& \epsilon_{S1}^{sp}(\omega,\omega',\beta_h,\beta_c) = \frac{3 \beta_c \beta_h^2}{4 \omega^2 \omega'^2 \Bigg(\beta_c \ln \Big[\frac{\cosh (\frac{\beta_h \omega'}{2})}{\cosh (\frac{\omega \beta_h}{2})} \Big] +\beta_h \ln \Big[ \frac{\cosh (\frac{\omega \beta_c}{2})}{\cosh (\frac{\beta_c \omega'}{2})}\Big] \Bigg)^2} \Bigg(\omega'^2 \Big(\tanh (\frac{\omega \beta_c}{2})-\tanh (\frac{\omega \beta_h}{2})\Big)+\omega^2 \Big(\tanh (\frac{\beta_h \omega'}{2})-\tanh (\frac{\beta_c \omega'}{2})\Big)\Bigg) \nonumber\\
& \times \Bigg(2 \ln \Big[\frac{\cosh (\frac{\omega \beta_c}{2})}{\cosh (\frac{\beta_c \omega'}{2})}\Big] -\omega \beta_c \tanh (\frac{\omega \beta_c}{2})+\beta_c \omega' \tanh (\frac{\beta_h \omega'}{2})\Bigg) \nonumber\\
& -\frac{3 \beta_c \beta_h \Bigg(-2 \omega \tanh (\frac{\beta_c \omega'}{2})-\beta_c \omega'^2 \text{sech}^2(\frac{\omega \beta_c}{2})+\omega \Big(\beta_h \omega'+\sinh (\beta_h \omega')\Big) \text{sech}^2(\frac{\beta_h \omega'}{2})\Bigg)}{4 \omega  \omega'^2 \Bigg(\beta_c \ln \Big[\frac{\cosh (\frac{\beta_h \omega'}{2})}{\cosh (\frac{\omega \beta_h}{2})} \Big] +\beta_h \ln \Big[\frac{\cosh (\frac{\omega \beta_c}{2})}{\cosh (\frac{\beta_c \omega'}{2})}\Big]\Bigg)}
\end{align}

\section{COP of classical anharmonic Otto refrigerator} \label{A2}

Under the framework of endoreversible thermodynamics~\citep{endo} or the notion of local equilibrium, we consider a classical anharmonic oscillator or a single Brownian particle in an anharmonic trap as working medium of the 4-step Otto refrigerator. The concept of endoreversible refrigeration emerges from the irreversible nature of the classical world with a system taking much time to attain an equilibrium with the bath. While the temperature of the system changes gradually maintaining a local equilibrium with the bath at all times, it is not in global equilibrium though from the perspective of the bath. The performance of classical Otto engines using harmonic oscillator is analysed in~\citep{Deffner}.

The energy for classical anharmonic oscillator with quartic perturbation has the form given by Eq.(1) (main text), where position, $x$ and momentum, $p$ are not operators. As $\lambda$ has dimension of (frequency)$^3$,  here too $\frac{\lambda}{\omega_0^3} \in (0,1)$ is a dimensionless variable. From phase-space analysis, the partition function corresponding to the classical anharmonic oscillator up to first order in $\lambda$ is given by, 
\begin{equation} 
Z^{ao}_{classical}(\beta,\omega) = \frac{1}{\beta \omega} + \frac{3}{\beta^2 \omega^3} \lambda
\end{equation}
Using this expression, we obtain entropy and mean energy of the oscillator respectively, as (considering Boltzmann constant, $k_B$=1),
\begin{align}
& S^{ao}_{classical}(\beta,\omega)= 1-\log (\beta \omega) -\frac{6}{\beta \omega^4} \lambda \\
\mbox{and}~~~~ & E^{ao}_{classical}(\beta,\omega) = \frac{1}{\beta} - \frac{3}{\beta^2 \omega^4} \lambda
\end{align}

\begin{figure}[!ht]
\includegraphics[width=.3\linewidth]{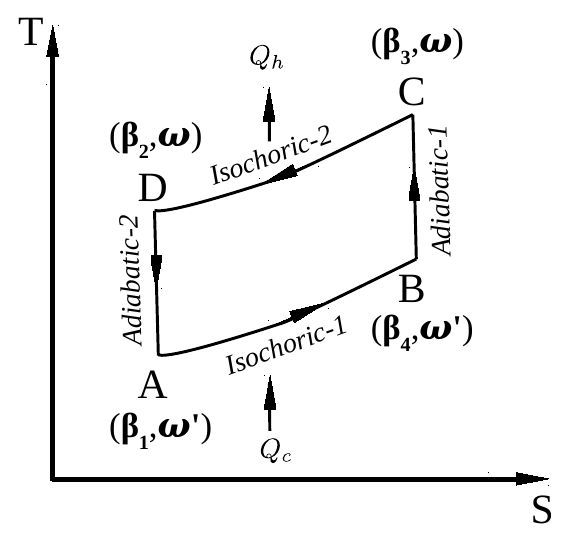}
\caption{\footnotesize Temperature-Entropy (T-S) diagram of classical endoreversible Otto cycle}
\label{cor}
\end{figure}

Fig.\ref{cor} schematically describes the process of refrigeration through classical endoreversible Otto cycle. It consists of two isochoric and two adiabatic processes. But unlike the quantum isochoric process, which eventually behaves as classical isothermal process, the classical isochoric process involves changes in both pressure and temperature, thus consisting of four different temperatures at four end points. During {\it Isochoric-1 (A$\rightarrow$ B)} process, the system remains in local equilibrium with the cold bath but never attains global equilibrium with the bath. Therefore, the inverse temperature of the system changes slowly from $\beta_1$ to $\beta_4$ where $\beta_c \geq \beta_1 > \beta_4$ ($\beta_c$ being the temperature of the cold reservoir). While doing no work (i.e. frequency pertaining to the Hamiltonian remains $\omega'$), the system absorbs heat from the cold reservoir as follows,
\begin{equation}
Q'_c = E^{ao}_{classical}(\beta_4,\omega')-E^{ao}_{classical}(\beta_1,\omega') 
\end{equation}
During {\it Adiabatic-1 (B$\rightarrow$ C)} process, the system does not exchange heat with the reservoir, hence the entropy remains fixed, i.e. $S^{ao}_{classical}(\beta_4,\omega') = S^{ao}_{classical}(\beta_3,\omega)$ while the frequency corresponding to the Hamiltonian changes from $\omega'$ to $\omega$. Therefore, work done on the system is given by,
\begin{equation}
W'_{comp} = E^{ao}_{classical}(\beta_3,\omega) - E^{ao}_{classical}(\beta_4,\omega')
\end{equation}
Next in the {\it Isochoric-2 (C$\rightarrow$ D)} process, the system remaining in local equilibrium with the hot reservoir (at $\beta_h$), slowly changes its inverse temperature from $\beta_3$ to $\beta_2$, where $\beta_h \leq \beta_3 < \beta_2$. As the Hamiltonian or corresponding frequency is kept fixed at $\omega$, the system rejects heat to the hot bath by an amount,
\begin{equation}
Q'_h = E^{ao}_{classical}(\beta_3,\omega)-E^{ao}_{classical}(\beta_2,\omega) 
\end{equation}
Finally, during {\it Adiabatic-2 (D$\rightarrow$ A)} process, the entropy remains fixed, i.e. $S^{ao}_{classical}(\beta_2,\omega) = S^{ao}_{classical}(\beta_1,\omega')$ and the frequency of the oscillator changes from $\omega$ to $\omega'$. Thus the work done by the system is,
\begin{equation}
W'_{exp} = E^{ao}_{classical}(\beta_2,\omega) - E^{ao}_{classical}(\beta_1,\omega')
\end{equation}

Now, we can compute the co-efficient of performance (COP) of the classical anharmonic Otto refrigerator as,
\begin{align}
\epsilon^{ao}_{O'} &=\frac{Q'_c}{|W'|}=\frac{Q'_c}{|Q'_h - Q'_c|} \nonumber\\
&= \frac{E^{ao}_{classical}(\beta_4,\omega')-E^{ao}_{classical}(\beta_1,\omega')}{(E^{ao}_{classical}(\beta_3,\omega)-E^{ao}_{classical}(\beta_2,\omega))-(E^{ao}_{classical}(\beta_4,\omega')-E^{ao}_{classical}(\beta_1,\omega'))} \nonumber\\
&=\frac{\omega'}{\omega-\omega'} + \frac{3\lambda}{\beta_1 \beta_2 \beta_3 \beta_4} \frac{(\beta_1 - \beta_4)(\beta_2 - \beta_3)}{(\beta_1 \beta_2 \beta_3 - \beta_2 \beta_3 \beta_4 + \beta_3 \beta_4 \beta_1 - \beta_4 \beta_1 \beta_2)^2} \Big( \frac{1}{\beta_2 \omega^4} + \frac{1}{\beta_3 \omega^4} - \frac{1}{\beta_4 \omega'^4} -\frac{1}{\beta_1 \omega'^4}\Big)
\end{align}

The quantity $(\frac{1}{\beta_2 \omega^4} + \frac{1}{\beta_3 \omega^4} - \frac{1}{\beta_4 \omega'^4} -\frac{1}{\beta_1 \omega'^4})$ becomes positive or negative depending on the  conditions: ($\frac{\beta_1}{\beta_2} > \frac{\omega^4}{\omega'^4}$, or  $\frac{\beta_1}{\beta_2} < \frac{\omega^4}{\omega'^4}$), and ($\frac{\beta_4}{\beta_3} > \frac{\omega^4}{\omega'^4}$, or $\frac{\beta_4}{\beta_3} < \frac{\omega^4}{\omega'^4}$). Therefore, anharmonicity may either increase or reduce the COP of the classical refrigerator.  In Fig.\ref{co} we present an example of the
choice of parameters for which the COP for the classical Otto refrigerator decreases with
anharmonicity, whereas the corresponding quantum COP increases  w.r.t. $\frac{\lambda}{\omega_0^3}$.
\begin{figure}[!ht]
\includegraphics[width=.3\linewidth]{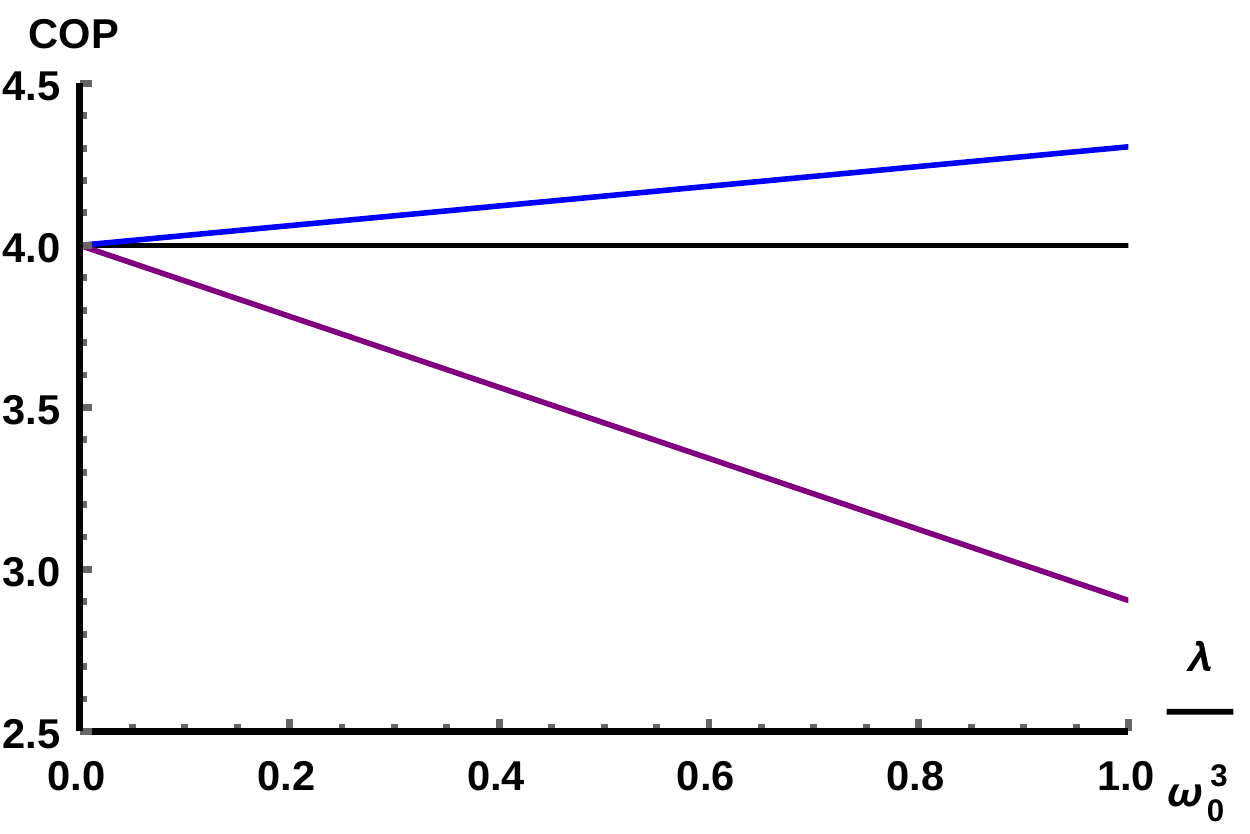}
\caption{\footnotesize (Color Online) COP of Otto refrigerator versus anharmonicity with parameters, $\omega$=5,  $\omega'$=4, $\beta_h=\beta_3 = \frac{1}{2}$, $\beta_c=\beta_1 =1$, $\beta_2 = \frac{3}{5}$, $\beta_4 = \frac{4}{5}$ and $\omega_0 = \sqrt[3]{0.6}$. The upper(blue), middle(black) and lower(purple) st. lines imply COP corresponding to quantum anharmonic oscillator, harmonic oscillator and classical anharmonic oscillator respectively.}
\label{co}
\end{figure}
The plot shows that for such a choice of parameters  though there is a decrease of COP by considerable amount with the strength of anharmonicity in case of the classical anharmonic oscillator, the quantum anharmonic oscillator provides improvement compared to the harmonic oscillator (classical or quantum).
\end{widetext}

\end{document}